Running head: CREATIVE IDEA ILL-DEFINED MENTAL REPRESENTATIONS

The Form of a 'Half-baked' Creative Idea:

Empirical Explorations into the Structure of Ill-defined Mental Representations

Victoria S. Scotney[a,1], Jasmine Schwartz[a], Nicole Carbert[a,2], Adam Saab[a], & Liane Gabora[a,3]

[a]University of British Columbia, Department of Psychology

Irving K. Barber School of Arts and Sciences, Fipke Centre for Innovative Research

3247 University Way, Kelowna, BC, Canada, V1V 1V7

[1]Present address: Department of Psychological Sciences, Purdue University

703 Third Street, West Lafayette, IN 47907-2081 USA

[2]Present address: University of British Columbia, School of Population and Public Health

2206 East Mall, Vancouver, BC, Canada, V6T 1Z3

[3]Address correspondence to: Liane Gabora (liane.gabora@ubc.ca)

Abstract

Creative thought is conventionally believed to involve searching memory and generating multiple independent candidate ideas followed by selection and refinement of the most promising. Honing theory, which grew out of the quantum approach to describing how concepts interact, posits that what appears to be discrete, separate ideas are actually different projections of the same underlying mental representation, which can be described as a superposition state, and which may take different outward forms when reflected upon from different perspectives. As creative thought proceeds, this representation loses potentiality to be viewed from different perspectives and manifest as different outcomes. Honing theory yields different predictions from conventional theories about the mental representation of an idea midway through the creative process. These predictions were pitted against one another in two studies: one closed-ended and one open-ended. In the first study, participants were interrupted midway through solving an analogy problem and wrote down what they were thinking in terms of a solution. In the second, participants were instructed to create a painting that expressed their true essence and describe how they conceived of the painting. For both studies, naïve judges categorized these responses as supportive of either the conventional view or the honing theory view. The results of both studies were significantly more consistent with the predictions of honing theory. Some implications for creative cognition, and cognition in general, are discussed.

## 1. Introduction

Creative ideas often take time to mature before they reach their final form (Feinstein, 2006). This leads to the interesting question: what form does a creative idea take midway through the creative process? Theories of creativity have rarely addressed this question directly.

It is widely assumed that creative thought involves two phases: the generation of multiple discrete ideas, followed by the selection and development of the most promising (e.g., Engelmann & Gettys, 1985; Finke, Ward, & Smith, 1992; Simonton, 1999). This implies that the mental representation of the creative idea goes from encompassing a set of different possibilities to encompassing only one possibility that undergoes refinement. An alternative theory of creativity (Gabora, 2017), which grew out of mathematical models of concept combination, is that throughout the creative process there is only one mental representation of the creative idea but, due to its inherent ambiguity, it can manifest (i.e., actualize) as different outputs when viewed from different perspectives (like the different shadows cast by the same object when lit from different directions). This paper presents two studies carried out to investigate the form of creative ideas during the creative process. We begin by outlining currently predominant theories of creativity and what they imply about the form of an idea midway through the creative process. We then present an alternative theory of creativity, honing theory, that leads to a different conception of the form of a creative idea midway through the creative process.

### 1.1. Search and Select Theories of Creativity

We will refer to the bulk of existing theories of creativity as *Search and Select (SS)* theories, because searching and selecting from amongst possible candidates features prominently in many such theories, though not all such theories entail search, and not all entail selection. Nor are search and selection necessarily more prominent components of SS theories than other processes such as divergent thinking and concept combination. However, divergent thinking and concept combination are also central to the alternative theory outlined in the next section, with which SS theories will be contrasted. Thus, the term 'Search and Select' was chosen partly because it emphasizes what distinguishes them, and should not be interpreted as distilling their core elements.

An example of an SS theory of creativity, the *Blind Variation Selective Retention* (BVSR) theory (Campbell, 1960; Simonton, 1999, 2007, 2010, 2013), asserts that creative ideas come about through a trial-and-error combinatorial process involving *blind* generation of ideational *variants* followed by *selective retention* of the fittest for development into a finished product. The term 'combinatorial' is conceived of in the broadest sense, including both processes and procedures, and producing either ideas or responses. The variants are said to be 'blind' in the sense that their final utilities are unknown at the time of initial generation (in other words, the creator has no subjective certainty about whether they are a step in the direction of the final creative product). Potential idea variants are generated sequentially or simultaneously (Simonton, 2013). While some are ultimately retained, others may be discarded at earlier stages. Although idea variants may have elements in common, each is regarded as a structurally distinct and separate entity, with its own probability and utility. Another example of a SS theory is the Geneplore model (Finke, Ward, & Smith, 1992) according to which, the creative process consists of a generative phase, where an individual constructs mental representations called preinventive structures, and an exploratory phase where those structures are used to come up with creative ideas. Existing computational theories of creativity are also SS theories (Besold, Schorlemmer, & Smaill, (2015).

The assumption that the variants of a creative idea, or the constituents of which an unfinished creative idea is composed, can be treated as distinct and separate entities is widespread. It is also implied in the *Structure Mapping theory* (SM) of analogical thought (Forbus, Gentner, & Law, 1995; Gentner, 1983, 2010). Analogy making is a central and ubiquitous facet of human creativity thatn features prominently in lists of creative processes (Finke, 1996; Martindale, 1998; Welling, 2007), and has played an important role in creative breakthroughs, inventions, and scientific discoveries (Gentner, 1989; Gentner & Markman, 1997). Analogy making involves developing a better understanding of something that is not well understood, the *target*, by drawing upon something that is better understood, the *source*. A key principle of the structure mapping theory of analogy is the *systematicity principle,* according to which people prefer to connect structures composed of related predicates. The theory posits that analogy generation begins with finding all possible source-to-target content matches through a 'structurally blind' quick and dirty parallel search of longterm memory that emphasizes surface similarity (Gentner, 2010; Gentner & Forbus, 2011). Potential sources are treated as distinct, pre-existing entities located in memory that can in this initial stage be considered separately from each other and from the rest of the mental contents. A promising source is selected, and the source and target are aligned through identification of common relational structure referred to as candidate inferences (or one-to-one correspondences). These candidate inferences start to get combined into clusters (or kernels), which are eventually merged to give a single mapping between source and target (Gentner & Smith, 2012).

These theories have much to offer and have advanced our understanding of creativity. However, while quite different, they bear some similarities that are called into question here. They treat ideas like objects in the physical world that can be independently chosen, assessed, and manipulated. They posit an initial stage in which each candidate idea or contributor to the final idea can be treated as distinct and separate from other such candidates, and from the rest of the contents of the mind. While incorporating the combining of *remote associates* (Mednick, 1962)—whether they be distantly related ideas, or analogues—they treat the remote associate as if it can be lifted, with its pre-existing ensemble of features or properties, and brought to bear on the creative task. They do not explicitly incorporate the notion of 'potentiality,' and thus do not contain a mechanism whereby the *context* elicits a never-before-noticed property in the associate. In other words, they do provide a means by which a property not conventionally thought to be related to the associate is, for the first time, retrieved as part of the associate, enabling the creator to generate, on the fly, the *basis* for the remote association.

## 1.2. Honing Theory of Creativity

Our everyday experience in a world of objects that exist in distinct locations and have definite boundaries may make it difficult to wean ourselves from the intuition that concepts and ideas in the mind behave this way as well (Gabora, 2019a,b). A more 'ecological' view is that they function not as fixed representations or identifiers, but as bridges between mind and world that are sensitive to context, and actively participate in the generation of meaning (Gabora, Rosch, & Aerts, 2008). This view is central to the *Honing Theory of creativity* (HT) (Gabora, 2010, 2017, 2018, 2019a, 2019b), which grew out of a particular mathematical approach to concept interactions (Aerts, Sozzo, & Veloz, 2016; Aerts, Broekaert, Gabora, & Sozzo, 2016; Aerts, Gabora, & Sozzo, 2013; Aerts & Gabora, 2005a, 2005b; Busemeyer & Bruza, 2012; Busemeyer & Wang, 2015; Gabora & Aerts, 2002; Pothos, Busemeyer, Shiffrin, & Yearsley, 2017). HT posits that we know ideas only indirectly, by how they appear from particular perspectives or contexts, but each context unavoidably influences how the idea is experienced and understood. The context the creator calls upon at any given instant reflects the current structure of his or her self-organizing internal model of the world, or worldview (Gabora, 1998, 2017; Gabora & Aerts, 2009). A key principle, to be elaborated shortly, is that during creative thought, what may appear to be distinct and separate ideas is in fact different actualizations or projections of the same idea considered from different contexts over the course of a creative 'honing' process, akin to different shadows cast by an unseen object when light is shone on it from different directions.

The issue of whether the mental representation of a creative idea is more like a physical object subject to search and selection or more like an unseen entity casting projections is not merely pedantic, for the formal structure of the two is entirely different (Gabora, 2005; Gabora & Aerts, 2007). To explain why, it will be necessary to briefly introduce what has come to be called (somewhat unfortunately) the quantum approach to concepts (Aerts, Broekaert, Gabora, & Sozzo, 2016b; Aerts, Gabora, & Sozzo, 2013; Blutner, Pothos, & Bruza, 2013; Busemeyer & Wang, 2018;; Gabora, 2001; Gabora & Aerts, 2002; Pothos, Busemeyer, Shiffrin, & Yearsley, 2017). It is called this not because it has anything to do with quantum particles, but because it uses generalizations of mathematical structures originally developed for quantum mechanics. The motivation and rationale for this approach are provided elsewhere (Aerts & Gabora, 2005; Atmanspacher, Romer, & Wallach, 2006; Bruza, Busemeyer, & Gabora, 2009; Busemeyer & Bruza, 2012; Khrennikov, 2010). For now, we note that this research does not aim to reduce cognitive psychology to physics, nor is it merely an anology. Rather, much as was the case with other branches of mathematics such as geometry and complexity theory, structures originally developed for physical sciences were later found to have applications in other domains.

Although the quantum approach to cognition is not just an analogy, the light-and-shadow analogy hinted at earlier can be a helpful as an explanatory aid to introduce the approach. In (Gabora, 2019b), the gist of the new view of creative thought is conveyed by the photograph below of woodcuttings with light shining on them from three different directions, yielding three differently shaped shadows for each woodcutting: that of a G, an E, and a B (Fig. 1). Consider first the upper woodcutting alone. Though each of its shadow is different, they are all projections of the same underlying object. We could say that the woodcutting has the *potentiality* to *actualize* different ways, and to actualize in one of these ways requires an *observable* or *context,* in this case, light shining from a particular direction. We can view the state of the woodcutting when no light is shining on it as its state of full potentiality. Similarly, while it is tempting to assume that a bout of creative thought entails the generation of multiple distinct, separate ideas, there may be a single underlying mental representation that, like the woodcutting, affords some degree of ambiguity in its interpretation. The fact that different sketches of a painting, or prototypes of an invention, take different forms when expressed in the physical world, doesn't mean they derive from different underlying ideas. Just as the shadows of the woodcutting are projections of one object, the sketches or prototypes may be different articulations of the same underlying idea looked at from different perspectives at different stages of a creative honing process. Midway through a creative thought process one may have an inkling of an idea but not yet know whether, or exactly how, it could work. Because it is 'half-baked', it may be more vulnerable to interpretation, meaning that it could appear quite different when looked at from a different perspective.

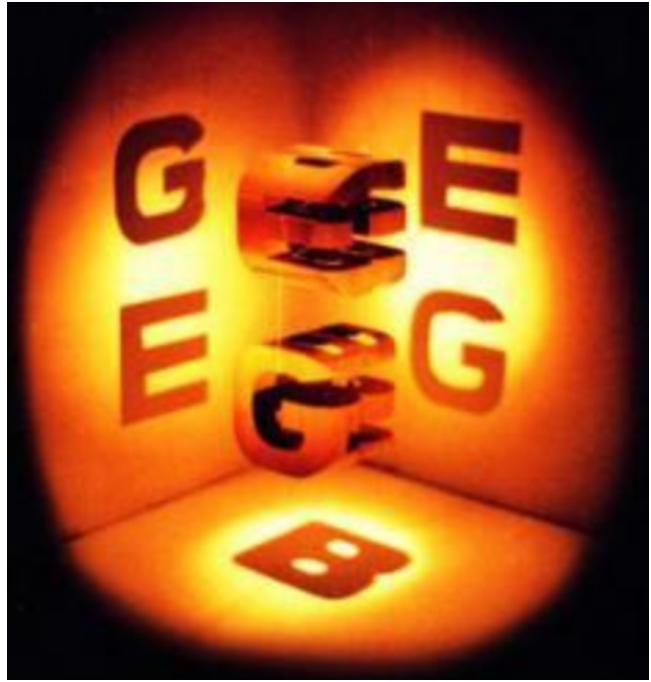

**Fig. 1**: Photograph of ambiguous woodcuttings taken from the front cover of '*Gödel, Escher, Bach: An Eternal Golden Braid' by* Douglas Hofstadter (1979). The top 'trip-let' (as he calls them) is not simply a rotated version of the one below it; it is a different shape. (Used with permission.)

Note that the two woodcuttings in Fig. 1 have different shapes, yet they yield the same three letters as shadows. To distinguish the shape of the woodcutting above from the woodcutting below would require that light be shown on them from still more angles, casting shadows that would not look like any actual letters of the alphabet. Similarly, the more complex one's unborn creative idea, the more honing steps that may be required to discern its underlying form and whittle it down as needed. Since it has the potential to manifest different ways, we can say that it is a *state of potentiality*. In the quantum approach, this kind of potentiality is described as a *superposition state*. More precisely, a superposition state is one that is indefinite with respect to some specific judgment. It is represented by a state vector in a complex Hilbert space, the dimensionality of which is determined by the number of unique elementary considerations. Possible outcomes (which play the role of observables in physics) constitute subspaces of this vector space. The outcomes are different ways in which the potentiality could be resolved.

Intuitively, the notion of a superposition state would appear to be well-suited to the description of mental representations at intermediate stages of a creative thought process, which are often imprecise and ambiguous. The ambiguity of a word or concept is often reduced through introduction of another word or concept that provides new contextual information, e.g., given the concept TREE you don't know whether it a deciduous tree (one with leaves) or a coniferous tree (one with needles and cones), but in the context CHRISTMAS, your mind likely settles on it being a coniferous tree. Each possible context may actualize the potentiality of the concept differently, and in the quantum approach these possible actualizations are represented by *basis states*. The actual, existing context is treated as an *observable* that determines how the concept changes in light of this context.[1] In the absence of any observable—i.e., when a concept is not being viewed from any particular context, or thought about at all—the concept is said to be in a *ground state*. In its ground state there are no properties associated with the concept, but also, there are no properties that are, a priori, excluded from it; thus, you could say it is a state of infinite potentiality. Conceptual change due to the impact of a context is modeled as a collapse event, which involves projection of the vector representing the concept in a superposition state to one of its basis states (Fig. 2).

---

[1] The context may affect the concept by altering the weights of certain properties. (For example, 'talks' and 'lives in a cage' are not considered properties of BIRD but they are considered properties of PET BIRD (Hampton, 1987); thus, the context PET is influencing the properties we ascribe to BIRD.) A context can also alter the typicalities of certain exemplars. (As a canonical example, GUPPY is not considered a typical exemplar of PET, nor of FISH, but it is considered a typical exemplar of PET FISH (Osherson & Smith, 1981).)

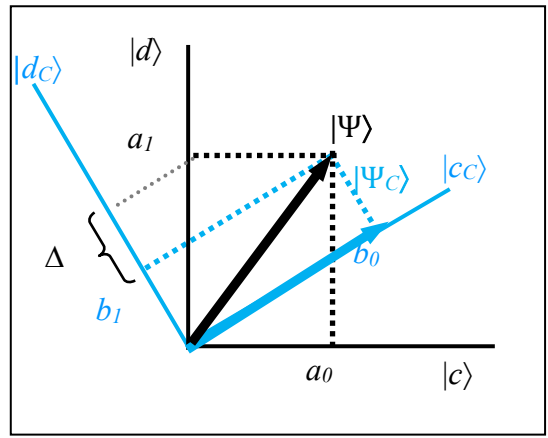

**Fig 2**: A graphical depiction of a vector $|\Psi\rangle$ representing the concept TREE is shown in black. In the default context, TREE may be more likely to collapse to projection vector $|d\rangle$ which represents DECIDUOUS TREE (tree with leaves) than to collapse to projection vector $|c\rangle$ which represents CONIFEROUS TREE (tree with needles and cones). This can be seen by the fact that subspace $a_0$ is smaller than subspace $a_1$; i.e., $a_0$ is closer to the xy origin than $a_1$. In the context CHRISTMAS, shown in blue, the concept TREE is likely to collapse to the orthogonal projection vector $|c_C\rangle$, representing CONIFEROUS TREE, as shown by the fact that $b_0$ is larger than $b_1$. (After collapse, the projected vector, $|\Psi_C\rangle$, is the same length as the original due to renormalization).

The concept of a ground state is consistent with the sparse, distributed, content-addressable structure of associative memory, which enables knowledge and memories that are relevant to the task at hand to come to mind naturally without search (Gabora, 2010, 2018). The notion of searching and selecting implies that the idea was in there all along and one just has to find it. However, memory does not work through a verbatim retrieval process; one does not so much retrieve an item as *reconstruct* it. Recalled items are unavoidably altered by what has been experienced since encoding, and spontaneously re-assembled in a way that relates to the task at hand (Brockmeier, 2010; McClelland, 2011). When remotely related items encoded in distributions of neural cell assemblies that only narrowly overlap are evoked simultaneously for the first time, the resulting widely sourced 'reconstruction' may well be difficult to parse and subject to interpretation.

The quantum approach provides a means to model some of the non-compositional ways in which people use concepts—famously said to be the biggest challenge facing cognitive science (Fodor, 1998)—by describing them in terms of effects such as superposition (described above), as well as entanglement[2] and interference[3] (Aerts, Gabora, & Sozzo, 2016; Aerts &

---

[2] Entanglement is a phenomenon first encountered in particle physics wherein the state of one entity cannot be described independently of the state of another, and any measurement performed on one influences the other. The presence of entanglement is detected by way of the violation of Bell's (1964) inequality, which proves that this mutual influence is not due to yet-to-be-identified 'hidden variables'. One example of entanglement with respect to concepts was provided in (Aerts, Gabora, & Sozzo, 2013). They conducted a study in which participants selected the best example of ANIMAL from amongst two options, e.g., BEAR and HORSE (A), or TIGER and CAT (A'), the best example of ACTS from amongst two options, e.g., GROWLS and WHINNIES (B), or SNORTS and MEOWS (B'), and in 'coincidence experiments', the best example of ANIMAL ACTS from permutations of the above (AB, A'B, AB', or A'B'). They demonstrated that the participants' responses violated a well-known variant of Bell's inequality (Clauser, Horne, Shimony, & Holt, 1969).

[3] Interacting waves generate patterns of constructive and destructive interference. Destructive interference is the annihilation of the crest of one wave by the trough of another, and constructive interference is the opposite, wherein crests and troughs are aligned, and amplify each other. The principle can be applied to concepts using data from studies involving conjunctions and disjunctions. For example, when people are asked to choose from amongst items

Sozzo, 2014; Busemeyer & Bruza, 2012). The approach can been applied not just to concept combination but also to more complex compounds of concepts such as decisions (e.g., Busemeyer, Wang, & Townsend, 2006; Yukalov & Sornette 2009) jokes (Gabora & Kitto, 2017), worldviews (Gabora & Aerts, 2009), and creativity (Gabora & Carbert, 2015). In creative thought, however, context does not necessarily reduce ambiguity, because (as noted above) the relationship between the original concept and the context may be remote, which may *introduce* ambiguity. For example, at the instant that Piero Gatti and colleagues first conceived of a beanbag chair, the context BEANBAG reduced ambiguity by constraining what kind of chair they would make, but it also introduced ambiguity, i.e., about whether such a chair would need legs, what kinds of beans to use, and so forth. As they thought it through they came up with different prototypes. However, in our view these prototypes were not independent entities; they were different manifestations of an evolving (where the term 'evolving' is used not as it is in biology but as it is in physics to refer to a dynamic process) mental representation of the 'beanbag chair' idea, which can be modeled as projections into different subspaces. Honing the idea is modeled as reiterated collapse, resulting in a change of state of the idea, which induces the creator to subject the idea to a new context, which brings about another collapse, and so such as COCONUT, ALMOND, and ACORN, which are good examples of FRUITS, good examples of VEGETABLES, and good examples of the disjunction FRUITS OR VEGETABLES, for some items the disjunction is chosen too often (overextension) while for others it is chosen too seldom (underextension) (Hampton, 1988). Plotting this pattern of overextension and underextension generated a distinctive pattern of constructive and destructive interference (Aerts, Broekaert, Gabora, & Sozzo, 2016).

forth, until the idea is sufficiently robust in the face of new contexts that it no longer undergoes change of state (Gabora, 2017).

An implication of HT that arises from the notion of ground state is that when an idea is not the subject of conscious attention—e.g., during incubation—it exists in a state of full potentiality in which no property is definitively present or absent. The context in which it comes to mind actualizes some properties and annihilates others. For a given property, there exists *some* possible context that could actualize (or annihilate) it, making it present (or absent). For example, although the property 'surrounded by water' would appear to be a defining property of the concept ISLAND, in the context KITCHEN—i.e., the concept combination KITCHEN ISLAND—that seemingly defining property is absent. On the other hand, other properties that are not usually true for ISLAND, such as 'found in a house' are true for KITCHEN ISLAND. It is in this sense that no mental content is a priori excluded; everything encoded in memory is fair game. That does not mean there are no constraints; in this case, for example, the context KITCHEN constrains which possible properties of ISLAND are present (or 'actualized') and which are not. Thus, unlikely properties are not ruled out *a priori,* they exist as possibilities, and conversely, seemingly defining properties are not 'ruled in', and it is this aspect of HT that makes it uniquely able it to explain how remote associates 'shake hands'.

## 1.3 Rationale for Current Studies

We have seen that while conventional theories predict that midway through a creative process one is searching and/or selecting from amongst multiple candidate ideas, HT predicts that one is modifying a single ambiguous representation that amalgamates components from different sources by viewing different context-specific articulations of it. This paper presents two studies that test between these theories. Study 1 presents an experiment with a close-ended analogy problem-solving task.[4] One might suppose that because analogy does not (in general) require that one come up with novel information so much as connect a source in memory to a structurally similar target, it is the creative process *most* likely to involve search and selection. Thus, if we can show that even analogy problem solving involves not search or selection amongst predefined alternatives but the honing of an ill-defined mental representation, we have compelling evidence for the hypothesis that honing figures prominently in this creative process.

Since it was not obvious that the state of half-baked creative ideas in one task would have anything in common with the state of half-baked creative ideas in another task, Study 2 presents an experiment conducted with an open-ended art-making task.[5]

## 2. Study 1

### 2.1. Background: Contrasting the Predictions of Structure Mapping versus Honing Theory

SM and HT give different predictions as to the state of the mind midway through analogy formation (as schematically illustrated in Fig. 3). The two theories give identical predictions regarding the form of completed analogies; it is only for incomplete analogies that they give different predictions. Therefore, we used an existing procedure that involves interrupting participants midway through a problem-solving session and asking them what they are thinking (Bowers, Farvolden, & Mermigis, 1995), and adapted it to an analogy problem-solving task.

| Structure Mapping | Honing Theory |
| --- | --- |

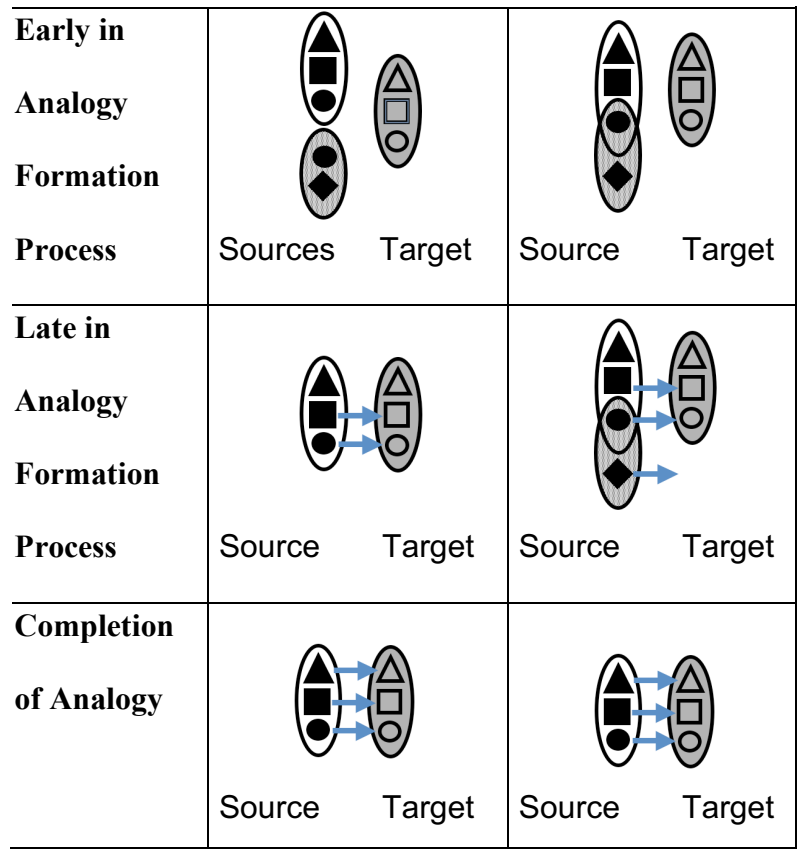

**Fig. 3.** A simplified illustration of the relevant differences between analogy solving by structure mapping versus honing theory. (See text for details.)

SM predicts that early in the analogy formation process, the analogist searches their mind for content matches from different sources. The matches are combined into structurally aligned, yet distinct, clusters (Fig. 3, top left). The analogist then selects the most structurally consistent cluster for mapping. Candidate inferences are projected sequentially (represented by two arrows), and the analogist evaluates the entire mapping after projecting all inferences. Upon reaching a correct solution, the problem solver has mapped and verified all three inferences (three arrows). Midway through analogy solving, participants are thought to be comparing different clusters and thus could write down multiple separate possibilities. In the case of structure mapping, there is no reason to expect that the incomplete solution will contain extra material.

According to HT, the mental representation is composed of multiple associated domains that share features (Fig. 3, top-right). By considering the idea from multiple perspectives, some correspondences will be mapped; unmappable ones will be deemed irrelevant for the current problem (two mapped arrows and one unmapped. By disambiguating the relevant from the irrelevant, the analogist knows what to exclude to reach the final well-defined solution (three arrows). Therefore, HT predicts that midway through the task, multiple possibilities may be given as an incomplete solution that will contain extraneous details. We hypothesize that midway through a creative task, the majority of participants will give responses that will be categorized as consistent with these predictions of HT.

## 2.2. Method

**2.2.1. Participants.** Eighty-five University of British Columbia undergraduates enrolled in psychology courses were recruited using an online study recruitment system (SONA). An additional six students participated as judges who evaluated the responses to the analogy task. (The sample size was simply the number of participants who signed up for the study.) They received course credit for their participation. All procedures were approved by the UBC Ethics Board.

**2.2.2. Materials.** The source and target used in this study were a pair of stories which are commonly used in the analogy literature (Gick & Holyoak, 1983). The source, titled 'The General', involves a fortress that cannot be captured if all soldiers come from the same direction but that can be successfully captured by dividing the army into small groups of soldiers that converge on the fortress from multiple directions (Appendix A).  The target, titled 'The Radiation Problem', involves finding a way to destroy a tumor without killing surrounding tissue (Appendix B).  The solution to The Radiation Problem is analogous to the solution to The General; the tumor is destroyed using multiple low-intensity X-rays from different directions.

**2.2.3. Procedure.** In the *exposure-to-source phase*, the participants were given five minutes to study The General. They were then asked to summarize the story as a test of their story comprehension. In the *problem-solving phase*, the participants were presented with the target, The Radiation Problem. They were given no indication that the story from phase one could help them solve the problem. Since pilot studies showed that the minimum time required to solve The Radiation Problem was two minutes, participants were interrupted after 100 seconds and told they had 20 seconds to write down whatever they were currently thinking in terms of a solution. Since two minutes was the minimum time to solve and not the average, the 100-second mark was deemed an appropriate time to interject. In a questionnaire distributed immediately afterward, they were asked whether they noticed a relation between The General and The Radiation Problem, and if so, at what point they noticed it.

**2.2.4. Judging.** Results of both phases were assessed by six judges who were naïve as to the theoretical rationale for the study. The story summaries produced in phase one were judged for comprehension on a three-point scale: poor, fair, or good.

Since we were interested in the nature of cognitive states midway through a creative process, participants who correctly solved the problem in the allotted time were removed from the analysis ($n = 34$). They were deemed to have correctly solved the problem if they included all three of the correspondences provided in Table 1.

Note that since we were not interested in the end product, but in the mental representation of an idea prior to arriving at a solution, we did not evaluate the creativity of participants' responses.

Table 1

*The Necessary Correspondences for a Complete Analogical Solution*

| The General (Source) | The Radiation Problem (Target) |
|---|---|
| 1. Soldiers separated into groups | 1. Multiple rays |
| 2. Each group is small in size | 2. Low intensity |
| 3. Groups converge from different directions | 3. Rays converge from different directions |

The judges were asked to categorize each of the remaining incomplete solutions as either structure mapping (SM) or honing theory (HT) according to the characteristics of each provided in Table 2. Judges were not asked to note whether criteria 1 and/or 2 were satisfied, and were only given SM and HT as options to choose between. (It was possible for an answer to contain elements predicted by both theories; for example it might contain multiple solutions separated by the word 'or' (indicative of SM), as well as extra information that would be relevant for related problems but that was not relevant for the one at hand (indicative of HT). However, using multiple judges all responses ended up categorized as one or the other.)

Table 2

*Characteristics Used to Judge Incomplete Analogy Solutions as Structure Mapping (SM) versus Honing Theory (HT)*

| Structure Mapping (SM) | Honing Theory (HT) |
|---|---|

| 1. If multiple solutions are given they are considered separate and distinct (for example, separated by the word 'or') | 1. If multiple solutions are given they are jumbled together |
|---|---|
| 2. Does not contain extra, irrelevant information | 2. Contains extra information that would be relevant for related problems but that is not relevant for this one |

A potential concern at this point is that an answer might contain extraneous information because it was elaborated following retrieval, rather than because the mental representation was ill-defined and consisted of multiple merged components. There is evidence that analogy making does in some cases involve adapting or elaborating the source to improve the match (Ross, 1987). SM assumes the analogy-solving process may involve generating and mapping *more* correspondences to extend the analogy after the three stages have been performed (Forbus, Gentner, & Law, 1995; Gentner & Markman, 1997), whereas HT argues that analogy-solving proceeds by *weeding out* non-correspondences that appear in the initial conception.

However, in the analogy used here, no adaptation or elaboration of the source was needed to obtain the complete and correct solution. In other words, the correct source (The General) could be used as is, without elaboration. Therefore, if extraneous information is present, we have good reason to believe that it was due to the presence of merged components.

An example of an answer that was categorized as SM is:

> *No idea. Don't know much about science. Maybe try to have a low-intensity ray that would sufficiently kill the tumor but not destroy healthy tissues.*

In this answer, one of correspondences has been mapped (correspondence 2: low intensity ray). Since the other two correspondences were not  mapped (multiple rays and different directions) the solution is incomplete. Since the answer contains no extraneous material, and because it provides no evidence that the participant's current conception of a solution consists of multiple items jumbled together in memory, it was classified as SM.

An example of an answer that was categorized as HT is:

> *The high intensity ray is necessary to kill the tumor so maybe shooting it in short successive bursts from different angles will kill the tumor without killing too much healthy tissue.*

In this incomplete solution, one of the correspondences has been mapped: different directions (correspondence 3). It was classified as HT because it includes irrelevant information (the notion of 'short successive bursts') activated by the target that is unnecessary to arrive at the correct solution.

## 2.3. Results

Analysis of Cohen's kappa between each pair of raters showed that one judge had poor agreement with all others on the ratings of story comprehension ($\kappa \leq 0.11$), and another judge had poor agreement with all others on the ratings of the incomplete solutions ($\kappa \leq 0.18$); therefore, the ratings from these judges were removed. An average of the kappa coefficient for each rater pair of the remaining five judges was calculated (Hallgren, 2012; Light, 1971) and showed fair agreement between the five judges for both ratings of story comprehension ($\kappa = 0.38$), and ratings of the incomplete solutions ($\kappa = 0.34$; Landis & Koch, 1977). Thirty-six participants responded that they saw a relation between the story and the problem, and 15 said they did not. Pearson's chi-squared test showed that there was no relationship between the categorization of a participants' story comprehension and the categorization of their incomplete solution, $\chi^2(2) = 0.64$, $p = .72$. There were two cells with expected frequencies less than five, but collapsing 'fair' and 'good' ratings showed a similar, nonsignificant result, $\chi^2(1) = 0.22$, $p = .63$. This indicates that the theory supported by a given response was not dependent on the level of the participant's story comprehension.

Each of the 51 incomplete solutions (those that remained after complete solutions were removed from the analysis) was classified as supportive of structure mapping if three or more judges judged it as type 'SM', and as supportive of honing theory if three or more judges judged it as type 'HT'. The odd number of judges meant that no cases were tied. 39 were classified as supportive of honing theory and 12 as supportive of structure mapping (Fig. 4). A one-sample chi-square test was significant, $\chi^2(1) = 14.29$, $p < .001$. Thus, the frequency count data more strongly support the predictions of Honing Theory than Structure Mapping.

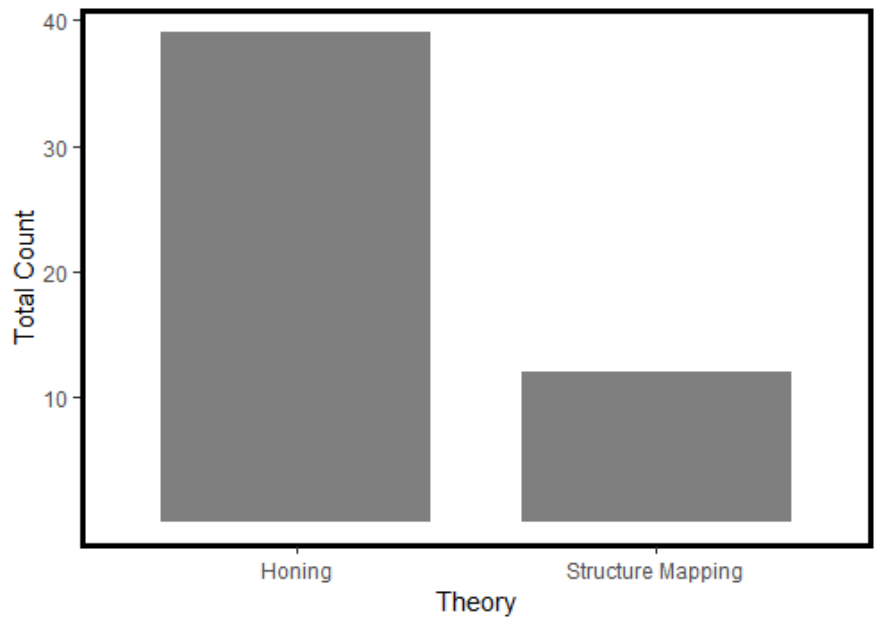

**Fig. 4.** Frequency of Solution Judgments for Honing Theory (n = 39) and Structure Mapping (n = 12).

A further analysis compared the mean number of judgments (out of a maximum of five,

the total number of judges) that supported each theory across all 51 responses. The mean number of SM judgments was 1.37 ($SD = 1.55$), and the mean number of HT judgments was 3.65 ($SD = 1.56$). A paired-sample t test showed that the difference was moderate and significant $t(50) = 5.23$, $p < .001$, $d = 0.66$. Thus, these data corroborate the above frequency count findings.

## 2.4. Discussion

Both the data from frequency counts and mean number of SM versus HT judgments are more consistent with the predictions of HT than SM, suggesting that HT more accurately describes the state of the mental representation of an analogy midway through the creative process. We note that different elements of the base analog were incorporated in different ways in different participants' answers, which shows that lack of a correct answer was not merely a problem of retrieval but of how to use the retrieved information. We reiterate that we were not assessing how creative the responses were, nor participants' success in carrying out the analogies, but rather the form of the mental representations involved.

We note that HT is not inconsistent with the notions of structural alignment and mapping, but merely SM's treatment of candidate structures as discrete and separate entities. The fact that a minority of participants' answers were judged as SS suggests that both theories capture some aspects of the underlying psychology, but that honing theory better captures the mental representation of an incomplete analogy.

We now turn to a study that investigated the state of half-baked creative ideas in an open-ended art-making task.

## 3. Study 2

## 3.1. Background

Study 1 provided an initial source of evidence that the mental representation of an idea midway through the creative process consists not of multiple, discrete possibilities, but a single structure with multiple merged components. However, analogy problem solving is a highly constrained task with a single correct solution, and one could argue that this is uncharacteristic of most creative tasks. Moreover, it has been argued that classic insight problems such as this divorce the task from the solver's real-world context because they do not require interaction with an object (Vallée-Tourangeau, Steffensen, Vallée-Tourangeau, and Sirota, 2016). The goal of Study 2 was to test this hypothesis using a less constrained, open-ended task.

Art-making is an 'ill-defined' creative task (Sawyer, 2018). Even if multiple artists are given the same prompt, they will reach vastly different 'solutions' (e.g., a painting they are satisfied with). The target state is not fixed as it was in Study 1. No matter how well-defined an artist's conception of the painting may become through the art-making process, the open-ended, subjective nature of the goal state makes art a viable domain for a less constrained creative task.

As with the analogy-solving task, SS theories and HT make different predictions about the mental representation of artists' conceptions of their work *midway through* an art-making task. The SS approach used by the conventional BVSR theory predicts that the artist is searching for and selecting from amongst multiple, separate, distinct ideas about their painting, selecting the best for refinement, and discarding the rest. They may overlap considerably in their components but they are not the same entity (much as two people both have a mouth, nose, and so forth, but they are different people). For example, Simonton treats Picasso's *Guernica* sketches as discrete idea variants which Picasso blindly creates and then chooses from to pursue further; these selections lead him towards the finished product in a nonmonotonic fashion but allow him to cleanly backtrack to earlier variants (Simonton, 2007). The artist's ideas, like the sketches of different parts of the painting, are considered separate from one another, lack a core vision guiding the process, and do not result in emergent self-understanding or internal transformation.

Honing theory predicts that if multiple idea components are contributing to the artwork, they are retrieved on the basis of activated attributes and in an amalgamated and indistinct form. Contrary to the SS approach, components are not isolated in the mind, and each new mental action changes, however subtly, the *mental representation* as a whole—even if there is no noticeable difference in the painting or any preliminary sketches—such that, after deciding to exclude the sun the artist has previously imagined, he or she advances to a conception of the painting that is, again, sunless, *but only in the context of having considered and rejected the sun component* (Gabora, 2005). We believe it is by this reasoning, not by backtracking, that additional idea components emerge out of earlier ones. Since the components are connected to the same representation, each should contain emergent properties of a core idea the artist has in mind. Honing such a multifaceted construct should also lead to greater self-understanding and internal transformation because the combination of concepts is novel.

We hypothesized that HT will more accurately predict the state of mental representations midway through the subjective painting task, i.e., ideas are predicted to be ill-defined, jumbled together, and to contain emergent properties of a common core idea that facilitate self-understanding and internal transformation in the majority of the artists' responses.

**3.2. Method**

**3.2.1. Participants.** The participants were undergraduate students enrolled in psychology courses at the University of British Columbia who received course credit for participating. There were two types of participants: 56 who created paintings and answered questionnaires, who will be referred to as *artists*, and six who judged the artists' answers to the questionnaires, who will be referred to as *judges.* All participants were naïve concerning the rationale for the study. As in the first study, they received course credit for their participation, and all procedures were approved by the UBC Ethics Board.

**3.2.2. Materials and Procedure for Artists.** Between two and eight artists participated at a time. They were seated at desks placed in a large circle, with each individual facing outwards such that they could not see each other's art. Each artist's desk had paintbrushes of two different sizes and an ice cube tray containing seven different colors of acrylic paint (pink, white, yellow, green, brown, blue, and red). Each desk also had a glass of water and two pieces of paper towel for cleaning the paintbrushes, and a plastic plate for mixing colors. Additionally, on the desk were a set of ten Crayola washable watercolor paints in plastic jars, pencil crayon pastels, chalk pastels, oil pastels, a piece of watercolor paper, and a piece of paper for acrylic paints.

Each artist was also provided with a handout that said: 'Create a painting that expresses yourself in any style that appeals to you'. They were told that their artwork would not be judged in any way, nor photographed, and that only their responses to questions about the art-making process would be analyzed. They were told that the study was concerned with peoples' perceptions of the creative process of art-making. They had the option to pick up their painting at the end of the semester; otherwise, it was destroyed after one year.

Before beginning to paint, participants were given a form that asked, 'What are you thinking in terms of what your painting will look like? Write down your thoughts about your painting in as much detail as you can'. They were told to start painting as soon as they finished writing. After 15 minutes, they were interrupted and given another form that said, 'Write down your thoughts about the painting in as much detail as you can'. After they finished responding, they were asked to continue painting until they were finished or the time was up, whichever came first. At this point they were given a final form with 15 questions that said, 'Were all of your ideas for your painting distinct and separate ideas?' In a pilot study with 15 participants, there was an additional question: 'Do you think being interrupted affected how you carried out your painting?' Since all of the participants answered 'no' to this question it was omitted thereafter. (Had any of them answered 'yes' their data would have been omitted from the analysis and we would have continued asking this question of subsequent participants.) All participants completed the procedure within ninety minutes.

Wince we were interested in the states of mental representations, not in the artistic merit of the paintings, we did not use Creativity Self-Assessments in this study.

**3.2.3. Judging.** As in the analogy task, naïve judges assessed the artists' questionnaire responses. The judges did not see or evaluate any of the artworks; all they saw were the artists' questionnaire responses. Each judge was provided with a set of handouts. One handout included a set of criteria to classify artists' answers as indicative of one theory or the other (Appendix C), while a second handout included examples of the kinds of responses they may see, and how they should be coded. The judging criteria handout also contained a summary of differences between the two theories, provided in Table 3. The judges were told that a flat-out 'yes' to 'Were all of your ideas for your painting distinct and separate ideas?' was indicative of S and a flat-out 'no' was indicative of H. The judges were also told that for the question 'Was there ever a state in your art-making process where your conception of the painting felt "half-baked"?' a flat-out 'yes' was indicative of H, and a flat-out 'no' was indicative of S.

The judges were given the 56 questionnaires and a form on which to classify the artists' answers as indicative of either H or S. We told them, 'If you are not sure just go with your gut feeling'.

Table 3

*Characteristics Used to Judge Responses as Search and Select (S) versus Honing Theory (H)*

|  | **S** | **H** |
|---|---|---|
| If multiple ideas are given, they are | Distinct (e.g., complete ideas separated by 'or') | Jumbled together (e.g., idea fragments spliced together) |
| Ideas are | Well-defined; need to be tweaked/mutated and selected amongst | Ill-defined; need to be made concrete; later elements emerge from earlier ones |
| Common core to ideas? | Never | Yes, or sometimes |
| Emergent properties of painting? | No | Yes |
| Emergent self-understanding? | No | Yes |
| Emphasis | External product | Internal transformation |

### 3.3. Results

Agreement between the judges was calculated using an average kappa value from all the rater pairs, as in Study 1 (Hallgren, 2012; Light, 1971). For several questions there was poor agreement between raters, indicating that the judges may have had difficulty distinguishing between the predictions of the two theories. On the basis of the judges' level of agreement and the clarity of the questions, two questions were selected for analysis. The first was referred to as Question 1: 'Were all of your ideas for your painting distinct and separate ideas?' It had fair inter-judge agreement ($\kappa = 0.29$, $n = 55$; Landis & Koch, 1977). An example of a response to this question that was categorized as S is: 'Yes, they all required their own time and dedication to come together as one piece'. Another example is: 'Yes, I had several ideas which could have formulated into a drawing if I had chosen to pursue any one. I'm not sure why I chose the sunset'. An example from the same question that was coded as H is: 'No, they were about me. I put on a lot of masks and hide behind colours'. A second example is: 'No, the ideas flowed off of each other… filled in spaces that needed to be filled… if I liked one colour, I'd use it more often in another area of the page'. The second was referred to as Question 2: 'Was there ever a state in your art-making process where your conception of the painting felt "half-baked"?' It had fair agreement ($\kappa = .35$, $n = 47$). An example of a response to this question that was categorized as H is: 'Yes, initially when the sky was not painted, the painting felt "empty". This inspired the idea to paint the background'. An example coded as S is: 'Nope not really! I had a pretty good idea of what I wanted to paint'. Not all artists answered every question, and there were a few instances where a judge missed coding a response. One judge had poor agreement with the other judges, and removing this judge improved reliability for all but one question, so this judge was removed from the analysis. This improved agreement for Question 1 ($\kappa = 0.33$; fair agreement) and Question 2 ($\kappa = .51$; moderate agreement). Kappa values for all questions are displayed in Table 4.

Table 4

*Light's Kappa for Six Judges ($\kappa_{6judges}$) and Five Judges ($\kappa_{5judges}$). Analyzed Questions in Bold.*

| Question | $\kappa_{6judges}(n_{artists})$ | $\kappa_{5judges}(n_{artists})$ |
| --- | --- | --- |
| Pre-painting | | |
| 1. What are you thinking in terms of what your painting will | 0.17 (56) | 0.20 (56) |
| look like? | | |
| 2. Write down your thoughts about the painting in as much detail as you can. | 0.06 (55) | 0.08 (55) |
| Midway | | |
| 1. Write down your thoughts about the painting in as much detail as you can. | 0.08 (29) | 0.09 (29) |
| Post-painting | | |
| 1. Write down your thoughts about the painting in as much detail as you can. | 0.09 (53) | 0.11 (54) |
| 2. When you read the instructions, what ideas came to mind? | 0.11 (52) | 0.16 (52) |
| 3. What best describes the state of your mind when you first saw the directions? | -0.02 (38) | 0.02 (38) |
| 4. Try to describe how you came up with what you painted. | 0.10 (56) | 0.14 (56) |
| 5. Was there a point when you started to feel like your painting was coming together? If so, please explain. | 0.03 (55) | 0.03 (55) |
| 6. What did if feel like when your painting started coming together? | -0.02 (43) | -0.00 (43) |
| 7. *Did* your painting come together? | 0.09 (55) | 0.14 (56) |
| 8. Did your ideas for your painting all share the same core idea? | 0.21 (55) | 0.15 (56) |
| **9. Were all of your ideas for your painting distinct and separate ideas? (Question 1)** | 0.29 (55) | 0.33 (55) |
| 10. Do you think that you had complete and whole ideas that | 0.15 (55) | 0.28 (56) |
| you did not paint? | | |
| 11. Did you ever have multiple ideas about how to do a certain part of the painting and then make multiple sketches of that certain portion of the painting? (Yes or No) | 0.84 (51) | 0.88 (54) |
| 12. If you answered 'yes' to the previous question then please answer the following question: Did the creation of one sketch or a certain part of the painting affect how you went about creating the next sketch for the same part of the painting? | 0.03 (8) | 0.07 (15) |
| **13. Was there ever a state in your art-making process where your conception of the painting felt 'half-baked'? (Question 2)** | 0.35 (47) | 0.51 (47) |
| 14. Does your final painting look like you thought it would? Please explain your answer. | 0.10 (50) | 0.17 (50) |
| 15. Is there any sense in which you emerged from the painting with a better understanding of yourself? If so, please explain. | 0.12 (46) | 0.18 (47) |
| 16. Were you in any way transformed by the art-making process? If so, please explain. | 0.15 (27) | 0.17 (29) |

*Note.* Judge 2 missed coding a few responses, which is why $n_{artists}$ increases for some rows.

The artists' responses were classified as supporting HT if three or more of the five judges judged it as type H, and as supporting a conventional search and select theory if three or more judges judged it as S.

For Question 1, 39 of the artist responses were classified by the judges as supporting of theory H, and 16 were classified as supporting theory S (Fig. 5). A one-sample chi-square test revealed a statistically significant difference between the classifications for theory H and S, $\chi^2(1)$ = 9.62, $p$ = .002. For Question 2, 40 responses were coded in favor of theory H, while 12 were classified as supportive of theory S. A one-sample chi-square test showed that this difference was significant, $\chi^2(1)$ = 15.01, $p$ < .001.  Thus, the frequency counts are more consistent with the predictions of HT.

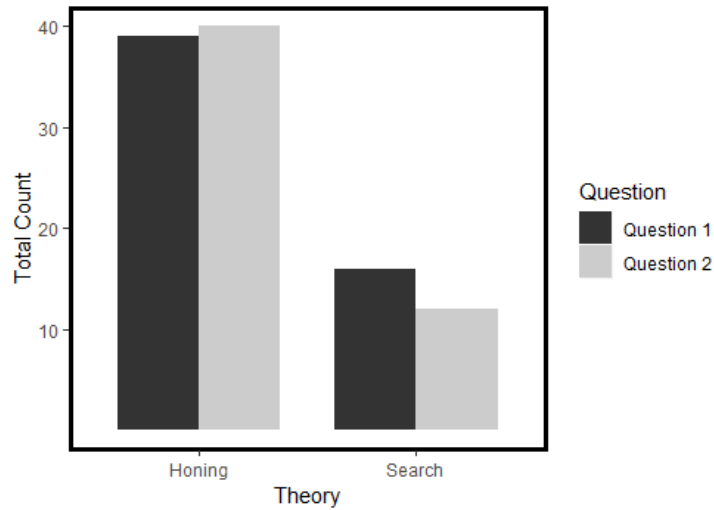

**Fig. 5**. Frequency of judgments of Honing Theory versus conventional Search and Select Theory for questions 1 and 2.

We then compared the mean number of judgments (out of a maximum of five, the total number of judges) across all responses that supported each theory. Taking the mean across all 55 responses for Question 1, the mean number of S judgments was 1.62 ($SD$ = 1.59), and the mean number of H judgments was 3.38 ($SD$ = 1.59). A paired-sample t test showed that the difference was significant and the effect size was moderate, $t(54)$ = 4.11, $p$ < .001, $d$ = 0.56. For Question 2, the mean number of S judgments was 1.53 ($SD$ = 1.74), and the mean number of H judgments was 3.36 ($SD$ = 1.72). A paired-sample t test showed that the difference was significant and had a moderate effect size, $t(51)$ = 3.82, $p$ < .001, $d$ = 0.55. Thus, these data corroborate the above frequency count findings.

## 3.4. Discussion

This study investigated the form of a creative idea midway through an open-ended art making task. As predicted by HT, the majority of participants considered their ideas about their painting ill-defined and inseparable from one another. This provides tentative support the hypothesis that midway through the creative process one is working with, not a collection of discrete candidates, but an ill-defined mental representation with the potentiality to manifest as different concrete outputs. Because of low agreement between judges, the questions regarding whether the art-making process facilitated emergent properties, self-understanding, and internal transformation remain unanswered.

## 4. General Discussion

This work contributes to our understanding of creative cognition by shedding light on the form ideas take during the creative process. We carried out two studies to investigate the mental representation of an unfinished idea. The first study showed that midway through an analogy problem the mental representation of the solution appears to take the form of, not a collection of independent solutions from amongst which one is selected for mapping, but rather, a conglomerate of potentially widely-sourced elements, as predicted by the honing theory of creativity (HT). The second study came to a similar conclusion using an open-ended art-making task. The results of both studies were in greater agreement with the predictions of HT over SS. They provide tentative support for the hypothesis that unborn ideas exist in a form that can be described as a superposition state.

There are many aspects of established theories of creativity—including the two mentioned in the introduction—that are not invalidated by these findings; indeed, elements of HT such as ground state and superposition state could be incorporated into BVSR or SM. One implication of these findings, however, and in particular their support for the 'incomplete idea as superposition state' hypothesis, is that they call into question the well-entrenched view that creativity involves two distinct forms of thought, alternately referred to as convergent versus divergent (e.g., Guilford 1967), generative versus exploratory (e.g., Finke, Ward, & Smith, 1992), associative versus analytic (e.g., Gabora, 2010), or executive versus generative (e.g., Ellamil, Dobson, Beeman, & Christoff, 2012).[6] As the creator starts to converge on a way of portraying the unborn idea that best communicates it to its intended audience, this may *appear* to entail a shift to a new and different kind of mental processing, i.e., from 'generating many' possible mental representations of the idea to 'refining one'. However, the real situation may be much as if one went from shining light on an object from different directions—generating multiple shadows—to shining light on it from similar directions—making minor tweaks on the 'same' shadow. There is just one entity casting shadows, and just one process: the process of shining a light and casting shadows. Our results are consistent with the proposal that in creative cognition there is just one representation of the unborn idea, modeled as a superposition state, and one process: nudging it closer to its final form by looking at it in different ways, which can be modeled by projecting it into different subspaces. The mental content of creative thought may be more vulnerable to manifesting as seemingly independent ideas because due to its ambiguity, there are more ways to think about (i.e., project) it. As you think through the creative idea, the considerations you subject it to (the contexts) start to converge, which could be modeled as the subspaces becoming more aligned, such that the rotation required to change from one evaluation basis to the next becomes smaller.[7] Thus, external manifestations of the creative process (i.e., design prototypes, or sketches made prior to the final painting) come to appear less like different ideas and more like variants on one increasingly well-defined idea. The mechanics of this shifting between what are conventionally construed as different modes of thought has been explored mathematically with a simple model of how the set of exemplars of a concept entertained is affected by a threshold of allowable deviation from the default context (Veloz, Gabora, Eyjolfson, & Aerts, 2011). The concept (HAT) is represented by a state vector that is a superposition of its possible exemplars (COWBOY HAT, etc.). Inserting data from a study in which participants were asked to rate the typicality of these different exemplars of a concept for different contexts (e.g., 'worn to be funny'), we showed how, by raising or lowering this threshold, the typicality of exemplars entertained increases or decreases. This suggests that the apparent transition from a divergent to a convergent mode of thought may be an artifact of viewing the idea from increasingly similar contexts. This view of creativity is consistent with research on artists and designers which suggest that creative ideation involves transforming a 'kernel idea' which goes from ill-defined to well-defined through an interaction between artist and artwork (Locher, 2010; Mace & Ward, 2002; Sawyer, 2018; Tovey, Porter, & Newman, 2003; Weisberg, 2004). Experiments on concept combination (summarized in Ward & Kolomytz, 2010) have shown that the more dissimilar the contributing concepts, the more original (though potentially less practical) the resulting idea. This suggests that the more dissimilar the 'raw ingredients' of an idea, the more honing it may require, and the more superposition states it must pass through to become viable.

---

[6] See (Sowden et al., 2014) for how the two forms of thought discussed in the creativity literature relate to other dual process theories. Indeed, the view proposed here also challenges dual process theories of cognition outside of the creativity literature.

---

[7] We might interpret this as reduced incompatibility of the considerations the idea is subjected to, such that the projectors are more nearly commutative (see Wang & Busemeyer, 2015).

This is the first set of studies we know of to use this 'interrupt midway through creative task' procedure, and as with all new experimental protocols, it has several limitations. The results could be affected by the introspective ability and eloquence of the participants. Mental states—and particularly, those concerning ideas that are not completely worked out—can be difficult to access and articulate, and experimental methods for studying introspection are underdeveloped (Ziegler & Weger, 2018). It could be that in the process of expressing their mental states verbally, SS mental states come out sounding like HT mental states, or *vice versa*. Since the results are quite strong and since this could work either for or against our hypothesis, we doubt that it would change the outcome, but it cannot be ruled out. Another limitation is that participants were disrupted at a particular point in time, and if they had been disrupted at a different point along the timeline, this could have influenced the relative degree of support obtained for the two theories. However, the window of opportunity to interrupt them was narrow, particularly for Study 1, since they required sufficient time to understand what the problem was and get an idea of how to approach it, but not so much time as to solve it.

There are also ways in which future studies could improve upon or develop further the methods used here. In future studies along the lines of Study 1, we suggest that providing judges with a 'neither theory' option would allow them greater flexibility. It could also be useful to incorporate a think-aloud procedure in addition to questionnaires, in order to capture the progression of ideas when working through a creative or problem-solving task (e.g., Pringle & Sowden, 2017). Both studies were somewhat limited by small sample size. Study 2 was limited by the difficulty the judges had in consistently rating the open-ended questions, which could be addressed by using more closed-ended questions, especially during the midway interruption. Future studies should improve the sample answers provided to train the judges, as we supplied seven of the sixteen questions, including the two chosen for analysis, with 'yes' or 'no' answers. We believe these two changes would increase agreement between the judges. As with Study 1, more categories could have been offered to the judges, such as 'neither' or 'both, to some degree'. In addition, in future studies, a broader range of tasks could be used. Future studies could also expand on the types of questions and carry out extensive pilot studies to assess their inter-rater reliability. All questions used were designed specifically for this study, and thus had not been previously tested for clarity or coding of responses.

Other possibilities for future studies involves testing for quantum effects in other ways. One such direction arises out of research in which decision making is modeled as collapse, i.e., projection of a state vector in a superposition state onto a subspace (e.g., Aerts & Aerts, 1995; Busemeyer & Bruza, 2012; Pothos & Busemeyer, 2009). This line of work has demonstrated the presence of asymmetries and order effects, as predicted by quantum theory (Aerts & Bianchi, 2017; Pothos, Busemeyer, & Trueblood, 2013). This may be relevant to creative cognition, since creativity involves making decisions. For example, prior to deciding what colour of paint to use next in a nonrepresentational piece of abstract art, it is not the case that the creator was 'lacking information' about what paint color came next, i.e., the decision did not involve a simple read-out from a pre-existing definite state. The decision was made on the fly given other constraints (such as, perhaps, the previously used paint colour, and/or the desired mood of the painting). By probing participants to make a decision or formulate an impression regarding some aspect of their creative work, and assessing the effect of that decision on subsequent cognitive processing along the lines of White, Pothos, and Busemeyer (2014), perhaps interrupting participants more than once, it may be possible to investigate whether creative cognition exhibits asymmetries and order effects.

We conclude by noting that much of human thought, not just creative thought, involves the resolution of ambiguity; thus, this work may have implications that extend beyond creativity. Nevertheless, we suspect that creative cognition may entail a unique 'secret sauce,' i.e., that it involves not just superposition but also, entanglement, and as discussed in Footnote 1, entangled states have been identified in cognition (through a nontrivial process). Studies are underway to assess whether concept combinations that meet the criterion for entangled states are perceived as more creative (Hinke & Gabora, in progress).

## 5. Acknowledgements

We thank Brian O'Connor for comments, and Apara Ranjan and Conner Gibbs for assistance with the manuscript. This work was supported by the Natural Sciences and Engineering Research Council of Canada (grant number 62R06523).

**Appendix A**

The following story, referred to as The General, was used as the analogy source in Study One: A small country was ruled from a strong fortress by a dictator. The fortress was situated in the middle of the country, surrounded by farms and villages. Many roads led to the fortress through the countryside. A rebel general vowed to capture the fortress. The general knew that an attack by his entire army would capture the fortress. He gathered his army at the head of one of the roads, ready to launch a full-scale direct attack. However, the general then learned that the dictator had planted mines on each of the roads. The mines were set so that small bodies of men could pass over them safely, since the dictator needed to move his troops and workers to and from the fortress. However, any large force would detonate the mines. Not only would this blow up the road, but it would also destroy many neighboring villages. It therefore seemed impossible to capture the fortress. However, the general devised a simple plan. He divided his army into small groups and dispatched each group to the head of a different road. When all was ready he gave the signal and each group marched down a different road. Each group continued down its road to the fortress so that the entire army arrived together at the fortress at the same time. In this way, the general captured the fortress and overthrew the dictator.

**Appendix B**

The following story, referred to as The Radiation Problem, was used as the analogy target in Study One:

Suppose you are a doctor faced with a patient who has a malignant tumor in his stomach. It is impossible to operate on the patient, but unless the tumor is destroyed the patient will die. There is a kind of ray that can be used to destroy the tumor. If the rays reach the tumor all at once at a sufficiently high intensity, the tumor will be destroyed. Unfortunately, at this intensity the healthy tissue that the rays pass through on the way to the tumor will also be destroyed. At lower intensities the rays are harmless to healthy tissue, but they will not affect the tumor either. What type of procedure might be used to destroy the tumor with the rays, and at the same time avoid destroying the healthy tissue?

**Appendix C**

**Judging Criteria:**

You will be trained how to read information about artists' art-making process and then put them into two categories: one that you think is indicative of one theory about the creative art-making process, and another that is indicative of another theory. You will first be given the identifying characteristics of Theory S or Theory H. You will practice on toy examples until your answers indicate that you understand the distinction between H and S, and until you yourself claim that you understand this distinction. Then you will classify the real responses as S or H.

**Identifying Characteristics of Theory S**

-        If multiple ideas are given, they are considered separate and distinct from one another; not hard to dis-entangle.

-        Does not contain extra ideas that would be relevant to other types of paintings or creative tasks but that are irrelevant to this particular painting.

-        Distinct possible ways of going about the task may be separated by words such as 'or' without anything to indicate these two ways are connected in the creator's mind.

-        The creative process involves searching one's mind for ways to go about the painting and selecting amongst these distinct possible outcomes.

-        No common core to possible painting outcomes.

-        No new emergent characteristics of painting come to light through process of resolving how initial idea for painting will be carried out.

**Identifying Characteristics of Theory H**

-        If multiple ideas are given, they are jumbled together; hard to dis-entangle.

-        Contains extra ideas that would be relevant to other types of creative tasks but that is not

relevant to creating this particular painting.

- Ill-defined or indistinct ideas; challenge is to make them concrete.

- Common core to different possible painting outcomes; core could be taken in different directions.

- Words indicative of H theory: vague, ill-defined, indistinct, potential.

- New emergent characteristics of painting, or new self-understanding come to light through process of resolving how initial idea for painting will be carried out; transformative.

Note: It is not necessary to meet all criteria in order to be properly classified as indicative of one theory of the other. Also, it is not necessary to distribute your classifications evenly; it is possible that 5 in a row might be indicative of one theory and not the other. In fact, it is completely possible that they all could be one theory and not the other. Please classify each answer as indicative of one theory or the other even if you are not sure of your answer.